\documentclass[proceedings, preprint]{rmaa}

\usepackage{paralist}

\usepackage{psfrag,color}

\title{Model of Reconnection of Weakly Stochastic Magnetic Field and its Implications} 

\author{
  A. Lazarian,\altaffilmark{1} 
  and E. Vishniac\altaffilmark{2}}

\altaffiltext{1}{Astonomy Department, University of Wisconsin-Madison, USA (lazarian@astro.wisc.edu).}

\altaffiltext{2}{Department of Physics \& Astronomy, McMaster University, Canada.}
\shortauthor{Lazarian \& Vishniac}
\shorttitle{Stochastic Reconnection}

\listofauthors{A. Lazarian \& E. Vishniac}
\indexauthor{Lazarian, A.}
\indexauthor{Vishniac, E.}

\abstract{We discuss the model of magnetic field reconnection in the presence of turbulence introduced
by us ten years ago. The model does not require any plasma effects to be involved in 
order to make the reconnection fast. In fact, it shows that the degree of magnetic field stochasticity 
controls the reconnection. The turbulence in the model is assumed to be subAlfvenic, with the magnetic field
only slightly perturbed. This ensures that the reconnection happens in generic astrophysical environments
and the model does not appeal to any unphysical concepts, similar to the turbulent magnetic diffusivity
concept, which is employed in the kinematic magnetic dynamo. The interest to that model has recently increased due to successful numerical testings of the model predictions. In view of this, we discuss implications of the model, including the first-order Fermi acceleration of cosmic rays, that the model naturally entails,
bursts of reconnection, that can be associated with Solar flares, as well as, removal of magnetic flux during
star-formation.}

\addkeyword{galaxies: magnetic fields- MHD}
\addkeyword{Sun: energetic particles}
\addkeyword{Sun: flares}
\addkeyword{Sun: magnetic fields}

\begin{document}
\maketitle

\section{Problem of Astrophysical Reconnection}
\label{sec:intro}
Plasma conductivity is high for most astrophysical circumstances.
This suggests that ``flux freezing'', where magnetic field lines move
with the local fluid elements, is a good approximation within
astrophysical magnetohydrodynamics (MHD). 

What happens when magnetic field lines intersect? This is the
central question of the theory of magnetic reconnection.  In fact, the
whole dynamics of magnetized fluids and the back-reaction of the
magnetic field depends on the answer.

\section{The Sweet-Parker Model versus Petschek Model}

The literature on magnetic reconnection is rich and vast (see e.g.,
Priest \& Forbes 2000 and references therein). We start by discussing a robust
scheme proposed by Sweet and Parker (Parker 1957; Sweet 1958).  In
this scheme oppositely directed magnetic fields are brought into
contact over a region of $L_x$ size (see Fig.~1). The diffusion of
magnetic field takes place over the vertical scale $\Delta$ which is
related to the Ohmic diffusivity by $\eta\approx V_r \Delta$, where
$V_r$ is the velocity at which magnetic field lines can get into
contact with each other and reconnect. Given some fixed $\eta$ one may
naively hope to obtain fast reconnection by decreasing $\Delta$.
However, this is not possible. An additional constraint posed by mass
conservation must be satisfied. The plasma initially entrained on the
magnetic field lines must be removed from the reconnection zone. In
the Sweet-Parker scheme this means a bulk outflow through a layer with
a thickness of $\Delta$.  In other words, the entrained mass must be
ejected, i.e., $\rho V_r L_x = \rho' V_A \Delta$, where it is assumed
that the outflow occurs at the Alfv\'en velocity.  Ignoring the
effects of compressibility, then $\rho=\rho'$ and the resulting reconnection
velocity allowed by Ohmic diffusivity and the mass constraint is
$V_r\approx V_A {\cal R}_L^{-1/2}$, where ${\cal R}_L^{-1/2}=(\eta/V_A
L_x)^{1/2}$ is the Lundquist number.  This is a very large number in
astrophysical contexts (as large as $10^{20}$ for the
Galaxy) so that the Sweet-Parker reconnection rate is negligible.

One of the ways to increase the reconnection rate is to change the global geometry
thereby reducing $L_x$. An example of the latter is the
suggestion by Petschek (1964) that reconnecting magnetic fields would
tend to form structures whose typical size in all directions is
determined by the resistivity (`X-point' reconnection).  This results
in a reconnection speed of order $V_A/\ln {\cal R}_L$.  However,
attempts to produce such structures in simulations of reconnection
have been disappointing (Biskamp 1984, 1986).  In numerical
simulations the X-point region  tends to collapse towards the
Sweet-Parker geometry as the Lundquist number becomes large (Biskamp
1996). A general review of astrophysical magnetic reconnection
theory can be found in Priest \& Forbes (2003). 

Petschek-type reconnection can survive in the presence of localized increase of effective resistivity,
which can happen due to various plasma effects. 
For instance, recent years have been marked by a substantial progress in simulations of collisionless reconnection (see Shay \& Drake 1998, Bhattacharjee et al. 2003, Drake et al. 2006a and references therein). This
work indicates that under some circumstances a kind of standing
whistler mode can stabilize an X-point reconnection region.  However,
these studies have not demonstrated the possibility of fast reconnection
for generic field geometries. They assume that there are no
bulk forces acting to produce a large scale current
sheet, and that the magnetized regions are convex, which minimizes
the energy required to spread the field lines. In addition, the requirements on the media being collisionless are rather strict, for instance, estimates in the literature (see Uzdensky 2006, Yamada et al. 2006) suggest that the $L_x$ should not exceed approximately 50 electron mean free paths. This makes the model not applicable to many astrophysical environments, e.g. to the interstellar medium.   

In any case, while
the researchers argue whether Hall MHD or fully kinetic description (Daughton 2006)  is necessary, one statement is definitely true. If magnetic reconnection
is only fast in collisionless environments, most of the MHD simulations, e.g. of interstellar medium, accretion disks, stars, where the
environment is collisional, are in error. We shall argue below that this radical conclusion may not be true and the reconnection is
also fast in most astrophysical collisional environments.

\section{Effects of Magnetic Stochasticity on Reconnection}

\subsection{Earlier Attempts}

The notion that magnetic field stochasticity might affect 
current sheet structures is not unprecedented.  In earlier
work Speiser (1970) showed that in collisionless plasmas the
electron collision time should be replaced with the time a
typical electron is retained in the current sheet.  Also
Jacobson \& Moses (1984) proposed that current diffusivity should be modified
to include diffusion of electrons across the mean field due
to small scale stochasticity.  These effects will usually be small
compared to effect of a broad outflow zone containing both
plasma and ejected shared magnetic flux.  Moreover,
while both of these effects
will affect reconnection rates, they are not sufficient to 
produce reconnection speeds comparable to the Alfv\'en speed
in most astrophysical environments. 

"Hyper-resistivity" (Strauss 1985, Bhattacharjee \& Hameiri 1986, Hameiri \& Bhrattacharjee 1987,
Diamond \& Malkov 2003) is a more subtle attempt to derive fast reconnection
from turbulence within the context of mean-field resistive MHD.  The form of the parallel
electric field can be derived from magnetic helicity conservation.  Integrating
by parts one obtains a term which looks like an effective resistivity
proportional to the magnetic helicity current.  There are several assumptions
implicit in this derivation.  The most important objection to this approach is that by
adopting a mean-field approximation one is already assuming some sort of
small-scale smearing effect, equivalent to fast reconnection.   Furthermore, the integration by parts involves assuming a large scale magnetic helicity flux through the boundaries of precisely the form required to drive fast reconnection.  Straus (1988)
partially circumvented the first problem by examining the effect of tearing
mode instabilities within current sheets.  However, the resulting reconnection
speed enhancement is roughly what one would expect based simply on the
broadening of the current sheets due to internal mixing.  This effect
does not allow us to evade the constraints on the global
plasma flow that lead to slow reconnection speeds, a point which
has been demonstrated numerically (see Matthaeus \& Lamkin 1985).

\subsection{Model in Lazarian \& Vishniac 99}

\begin{figure}[!t]
\includegraphics[width=\columnwidth]{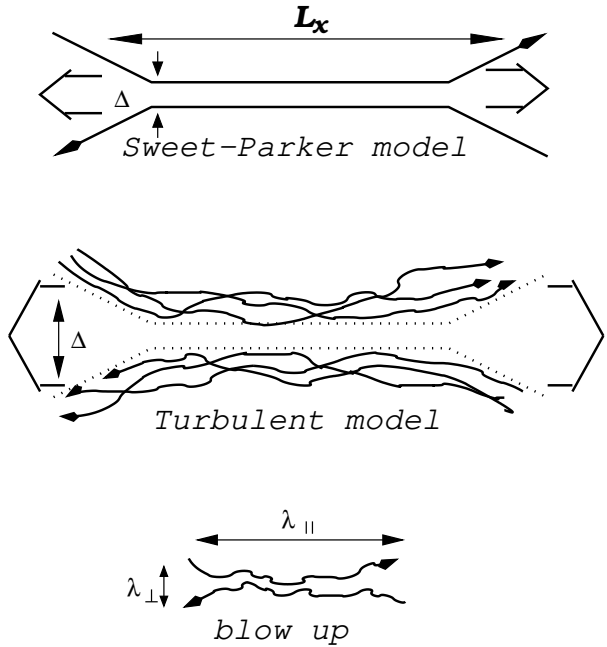}
\caption{{\it Upper plot}: 
Sweet-Parker model of reconnection. The outflow
is limited by a thin slot $\Delta$, which is determined by Ohmic 
diffusivity. The other scale is an astrophysical scale $L\gg \Delta$.
{\it Middle plot}: Turbulent reconnection model that accounts for the 
stochasticity
of magnetic field lines. The outflow is limited by the diffusion of
magnetic field lines, which depends on field line stochasticity.
{\it Low plot}: An individual small scale reconnection region. The
reconnection over small patches of magnetic field determines the local
reconnection rate. The global reconnection rate is substantially larger
as many independent patches come together.}
\label{recon}
\end{figure}

The model of reconnection proposed in Lazarian \& Vishniac (1999, henceforth LV99) is a natural generalization of the Sweet-Parker model (see Fig.~\ref{recon}). The
problem of the Sweet-Parker model is that the reconnection is negligibly slow for any realistic 
astrophysical conditions. However, astrophysical magnetic fields are generically turbulent.

LV99 consider the case in which there exists a large scale,
well-ordered magnetic field, of the kind that is normally used as
a starting point for discussions of reconnection.  
In addition, we expect that the field has some small scale `wandering' of
the field lines.  On any given scale the typical angle by which field
lines differ from their neighbors is $\phi\ll1$, and this angle persists
for a distance along the field lines $\lambda_{\|}$ with
a correlation distance $\lambda_{\perp}$ across field lines (see Fig.~\ref{recon}).

To do quantiative estimates one has to adopt the model of MHD turbulence. Such a model
for realistic compressible fluids may be constructed on the basis of the Goldreich-Sridhar (1995,
henceforth GS95) model of incompressible turbulence (see Cho \& Vishniac 2000, Maron \& Goldreich 2001,
Lithwick \& Goldrech 2001, Cho, Lazarian \& Vishniac 2002, Cho \& Lazarian 2002, 2003). The most
important in terms of magnetic reconnection are the properties of Alfvenic modes that determine field
wandering as shown in LV99.

The modification of the global constraint induced by mass conservation
 in the presence of
a stochastic magnetic field component 
is self-evident. Instead of being squeezed from a layer whose
width is determined by Ohmic diffusion, the plasma may diffuse 
through a much broader layer, $L_y\sim \langle y^2\rangle^{1/2}$ (see Fig.~\ref{recon}),
determined by the diffusion of magnetic field lines.  This suggests
an upper limit on the reconnection speed of 
$\sim V_A (\langle y^2\rangle^{1/2}/L_x)$. 
This will be the actual speed of reconnection if
the progress of reconnection in the current sheet does not
impose a smaller limit. The value of
$\langle y^2\rangle^{1/2}$ can be determined once a particular model
of turbulence is adopted, but it is obvious from the very beginning
that this value is determined by field wandering rather than Ohmic
diffusion as in the Sweet-Parker case.

What about limits on the speed of reconnection that arise from
considering the structure of the current sheet?
In the presence of a stochastic field component, magnetic reconnection
dissipates field lines not over their  entire length $\sim L_x$ but only over
a scale $\lambda_{\|}\ll L_x$ (see Fig.~\ref{recon}), which
is the scale over which magnetic field line deviates from its original
direction by the thickness of the Ohmic diffusion layer $\lambda_{\perp}^{-1}
\approx \eta/V_{rec, local}$. If the angle $\phi$ of field deviation
does not depend on the scale, the local
reconnection velocity would be $\sim V_A \phi$ and would not depend
on resistivity. In LV99 it is taken into account that $\phi$ does depend on scale.
Therefore the {\it local} 
reconnection rate $V_{rec, local}$ is given by the usual Sweet-Parker formula
but with $\lambda_{\|}$ instead of $L_x$, i.e. $V_{rec, local}\approx V_A 
(V_A\lambda_{\|}/\eta)^{-1/2}$.
It is obvious from Fig.~\ref{recon} that $\sim L_x/\lambda_{\|}$ magnetic field 
lines will undergo reconnection simultaneously (compared to a one by one
line reconnection process for
the Sweet-Parker scheme). Therefore the overall reconnection rate
may be as large as
$V_{rec, global}\approx V_A (L_x/\lambda_{\|})(V_A\lambda_{\|}/\eta)^{-1/2}$.
Whether or not this limit is important depends on
the value of $\lambda_{\|}$.  

The relevant values of $\lambda_{\|}$ and $\langle y^2\rangle^{1/2}$ 
depend on the magnetic field statistics. This
calculation was performed in LV99 using the GS95 model
of MHD turbulence providing 
the upper limit on the reconnection speed:
\begin{equation}
V_{r, up}=V_A \min\left[\left({L_x\over l}\right)^{\frac{1}{2}}
\left({l\over L_x}\right)^{\frac{1}{2}}\right]
\left({v_l\over V_A}\right)^{2},
\label{main}
\end{equation}
where $l$ and $v_l$ are the energy injection scale and
turbulent velocity at this scale respectively.
In LV99 other processes that can impede
reconnection were found to be less restrictive. For
instance, the tangle of reconnection field lines crossing the
current sheet will need to reconnect repeatedly before individual
flux elements can leave the current sheet behind.  The rate at which
this occurs can be estimated by assuming that it constitutes the
real bottleneck in reconnection events, and then analyzing each
flux element reconnection as part of a self-similar system of
such events.  This turns out not to impede the reconnection.  
As a result, LV99 
concludes that (\ref{main}) is not only an
upper limit, but is the best estimate of the speed of reconnection. 
The thick plasma outflows observed during the  2003 November 4 Coronal Mass Ejection reported
in Ciaravella \& Raymond (2008) are also consistent with the LV99 model.

\subsection{Generalizing the model for partially ionized gas}

 A partially ionized plasma fills a substantial volume within
our galaxy and the earlier stages of star formation take
place in a largely neutral medium. This motivates our study
of the effect of neutrals on reconnection.
The role of ion-neutral collisions is not trivial. On one hand, they
may truncate the turbulent cascade, reducing the small scale 
stochasticity and decreasing the reconnection
speed. On the other hand, the ability of neutrals to diffuse perpendicular
to magnetic field lines allows for a broader particle outflow 
and enhances reconnection rates.

Reconnection in partially ionized gases before the introduction of the LV99 model looked
hopelessly slow. For instance,  in Vishniac \& Lazarian (1999, henceforth VL99)
we studied the diffusion of neutrals away from the reconnection zone
assuming anti-parallel magnetic field lines
The ambipolar reconnection rates obtained in VL99, although large
compared with the Sweet-Parker model, are insufficient either
for fast dynamo models or for the ejection of magnetic flux prior
to star formation.  In fact, the increase in the reconnection speed
stemmed entirely from the
compression of ions in the current sheet, with the consequent enhancement of
both recombination and ohmic dissipation.  This effect is small
unless the reconnecting magnetic field lines are almost exactly
anti-parallel (VL99, see also Heitsch \& Zweibel 2003ab).  Any dynamically significant shared field component
will prevent noticeable plasma compression in the current
sheet, and lead to speeds practically indistinguishable from the standard
Sweet-Parker result.  Since generic reconnection regions will have
a shared field component of the same order as the reversing component,
the implication is that reconnection and ambipolar diffusion do not
change reconnection speed significantly.

Lazarian, Vishniac \& Cho (2004, henceforth LVC04) presented a model of turbulence in a partially ionized
gas. This model agrees well with numerical simulations available as the limiting case which
can be characterized by one fluid with a high Prandtl number (see Cho et al. 2002, 2003). 
Using this model LVC04 described field wandering, which is the core of the LV99 model of reconnection.
They showed that the magnetic reconnection proceeds fast both in diffuse interstellar and molecular cloud 
partially ionized gas.

\section{First order Fermi Acceleration Induced by Stochastic Reconnection}

In Sweet-Parker model the reconnection can accelerate charged particles, e.g. due
to the electric field in the reconnection region. However, the speed of Sweet-Parker 
reconnection is negligible for most astrophysical environments, thus the transfer of
energy from the magnetic field to particles is absolutely negligible, if the reconnection
follows the Sweet-Parker predictions.

It is interesting to notice that the first-order Fermi acceleration process
is intrinsic to the LV99 model of reconnection. To understand it
consider a particle entrained on a reconnected 
magnetic field line (see Fig.\ref{recon}). This particle
may bounce back and forth between magnetic mirrors formed by oppositely
directed magnetic fluxes moving towards each other with the velocity
$V_R$. Each of such bouncing will increase the energy of a particle
in a way consistent with the requirements of the first-order Fermi
process (de Gouveia Dal Pino \& Lazarian 2001, 2003, 2005, Lazarian 2005).

Another way of understanding the acceleration of energetic particles in the reconnection
process above is to take into account that the length of magnetic field lines is decreasing
during reconnection. As a result, the physical volume of the energetic particles entrained on
the field lines is shrinking. Thus, due to Louiville theorem, their momentum should increase to
preserve the constancy of the phase volume.  

An interesting property of this mechanism that potentially
can be used  to test the acceleration observationally is that the 
resulting spectrum of accelerated particles is different from that arising from a shock.
Gouveia Dal Pino \& Lazarian (2003, 2005) used this mechanism of particle acceleration to explain the synchrotron power-law spectrum arising from the flares of the
microquasar GRS 1915+105. Note, that the mechanism acts in the
Sweet-Parker scheme as well as in the scheme of turbulent reconnection.
However, in the former the rates of reconnection and therefore the
efficiency of acceleration are marginal in most cases.

\begin{figure}[!t]
\includegraphics[width=\columnwidth]{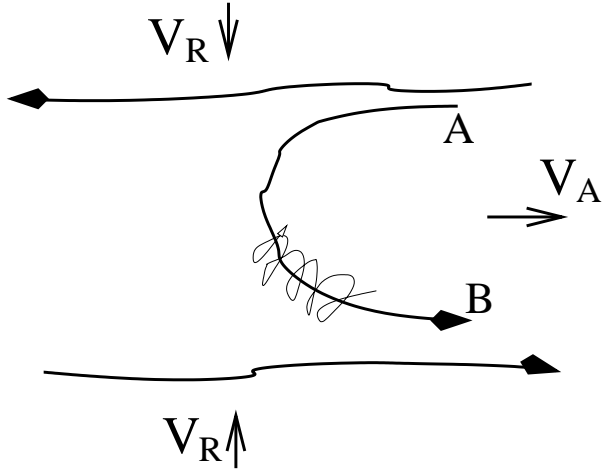}
\caption{  
Cosmic rays spiral about a reconnected magnetic
field line and bounce back at points A and B. The reconnected
regions move towards each other with the reconnection velocity
$V_R$. The advection of cosmic rays entrained on magnetic field
lines happens at the outflow velocity, which is in most cases
of the order of $V_A$. Bouncing at points A and B happens
because either of streaming instability or turbulence in the
reconnection region (from Lazarian 2005).}
\label{recon2}
\end{figure}

The mechanism
is similar to the acceleration mechanism that was proposed 
later by Drake et al. (2006b). Drake et al. (2006b) considered the acceleration of electrons and, similarly, to the Matthaeus, Ambrosiano \& Goldstein (1984), assumed that the acceleration happens within 2D contracting loops. While for LV99 model of reconnection the generic configuration of magnetic field are contracting  spirals this does not induce a radical difference between the processes. The difference in the spectrum of accelerated particles obtained by Drake et al. (2006b) and that obtained by us stems from the fact that in the former
paper the strong backreaction of the accelerated electrons was assumed. However, depending on the environment, this backreaction may vary, resulting in variations of the spectral slopes. Further studies of the acceleration process (see Fig.~\ref{recon2}) are necessary.

\section{Flares Induced by Stochastic Reconnection}

 If turbulence can drive reconnection, which in turn transforms magnetic energy into kinetic energy, then it seems appropriate to wonder if the process can be self-sustaining.  That is, given a very small level of ambient turbulence, how likely is it that reconnection will speed up as it progresses, without any further input from the surrounding medium?  A detailed examination of this is beyond the scope of current calculations, but we can clarify this with a simple physical model.

Let's consider a reconnection region of length $L$ and thickness $\Delta$.  The thickness is determined by the diffusion of field lines, which is in turn determined by the strength of the turbulence in the volume.  Reconnection will allow the magnetic field to relax, creating a bulk flow.  However, since stochastic reconnection is expected to proceed unevenly, with large variations in the current sheet, we can expect that some unknown fraction of this energy will be deposited inhomogeneously, generating waves and adding energy to the local turbulent cascade.  We take the plasma density to be approximately uniform so that the Alfv\'en speed and the magnetic field strength are interchangeable.  The nonlinear dissipation rate for waves is
\begin{equation}
\tau_{nonlinear}^{-1}\sim\max\left[ {k_\perp^2 v_{wave}^2\over k_\|V_A},k_\perp^2 v_{turb}\lambda_{\bot, turb}\right],
\end{equation}
where the first rate is the self-interaction rate for the waves and the second is the dissipation rate by the ambient turbulence (see Beresnyak \& Lazarian 2008ab)  The important point here is that $k_\perp$ for the waves falls somewhere in the inertial range of the strong turbulence.  Eddies at that wavenumber will disrupt the waves in one eddy turnover time, which is necessarily less than $L/V_A$.  The bulk of the wave energy will go into the tubulent cascade before escaping from the reconnection zone.  (This zone will radiate waves, for the same reason that turbulence in general radiates waves, but it will not significantly impact that energy budget of the reconnection region.)

We can therefore simplify our model for the energy budget in the reconnection zone by assuming that some fraction $\epsilon$ of the energy liberated by stochastic reconnection is fed into the local turbulent cascade.  The evolution of the  turbulent energy density per area is
\begin{equation}
{d\over dt}\left(\Delta v_{turb}^2\right)=\epsilon V_A^2 V_{rec}-v_{turb}^2\Delta {V_A\over L},
\end{equation}
where the loss term covers both the local dissipation of turbulent energy, and its advection out of the reconnection zone.  Since $V_{rec}\sim v_{turb}$  and $\Delta\sim L(v_{turb}/V_A)$,  we can rewrite this by defining ${\cal M}_A\equiv v_{turb}/V_A$ and $\tau\equiv L/V_A$ so that
\begin{equation}
{d\over d\tau}{\cal M}_A^3\approx \epsilon {\cal M}_A-{\cal M}^3.
\end{equation}
If $\epsilon$ is a constant then 
\begin{equation}
v_{turb}\approx V_A\epsilon^{1/2}\left[1-\left(1-{{\cal M}_0^2\over\epsilon}\right)e^{-2\tau/3}\right]^{1/2}.
\end{equation}
This implies that the reconnection rate rises to $\epsilon^{1/2}V_A$ is a time comparable to the ejection time from the reconnection region ($\sim L/V_A$).  Given that reconnection events in the solar corona seem to be episodic, with longer periods of quiescence, this implies that either $\epsilon$ is very small, for example - dependent on the ratio of the  thickness of the current sheet to $\Delta$, or is a steep function of ${\cal M}_A$.  If it scales as ${\cal M}_A$ to some power greater than two then initial conditions dominate the early time evolution.  

An alternative route by which stochastic reconnection might be self-sustaining would be in the context of a series of topological knots in the magnetic field, each of which is undergoing reconnection.  Now the problem is sensitive to geometry.  Let's assume that as each knot undergoes reconnection it releases a characteristic energy into a volume which has the same linear dimension as the distance to the next knot.  The density of the energy input into this volume is roughly $\epsilon V_A^2 v_{turb}/L$, where $\epsilon$ is the efficiency with which the magnetic energy is transformed into turbulent energy.  We have
\begin{equation}
\epsilon V_A^2{v_{turb}\over L}\sim {v'^3\over L_k}, 
\end{equation}
where $L_k$ is the distance between knots and $v'$ is the turbulent velocity created by the reconnection of the first knot.  This process will proceed explosively if $v'>v_{turb}$ or
\begin{equation}
V_A^2 L_k\epsilon> v_{turb}^2 L.
\end{equation}
This condition is almost trivial to fulfill.  The bulk motions created by reconnection will unavoidably generate significant turbulence as they interact with their surrounding, so $\epsilon$ should be of order unity.  Moreover the length of any current sheet should be at most comparable to the distance to the nearest distinct magnetic knot.  The implication is that each magnetic reconnection event will set off its neighbors, boosting their reconnection rates from $v_{turb}$, set by the environment, to $\epsilon^{1/2}V_A(L_k/L)^{1/2}$ (as long as this is less than $V_A$).  The process will take a time comparable to $L/v_{turb}$ to begin, but once initiated will propagate through the medium with a speed comparable to speed of reconnection in the individual knots.  In a more realistic situation,  the net effect will be a kind of  modified sandpile model for magnetic reconnection in the solar corona and chromosphere.  As the density of knots increases, and the energy available through magnetic reconnection increases, the chance of a successfully propagating reconnection front will increase.

\section{Stochastic Reconnection and Removal of Magnetic Flux from Molecular Clouds}

As we mentioned above, Sweet-Parker reconnection is too slow\footnote{A modification of the Sweet-Parker model to include ambipolar diffusion cannot generically induce much faster reconnection either (see Vishniac \& Lazarian 1999).}, while collisionless Petschek reconnection
is not applicable to molecular clouds. At the same time, Shu et al. (2006) showed that magnetic field 
is being removed from the star-forming cores faster than it is allowed by the standard ambipolar diffusion
scenario (see Tassis \& Mouschovias 2005). Shu et al. (2007) proposed a mechanism that required efficient
reconnection of magnetic loops referring to the ``hyperresistivity'' concept. As we discussed earlier,
this concept is not self-consistent and problematic at its core.
Could the reconnection be done by the mechanism we discuss in the paper?

As we discussed earlier, LVC04 considered magnetic reconnection in partially ionized gas and obtained fast reconnection rates that can be obtained there. Thus, it is suggestive that magnetic field can be removed this way from molecular clouds. Incidentally, this also means that the model of reconnection should be considered
for the transport of the angular momentum in protostellar disks (see Lazarian 2005).

\section{Discussion and Summary}

The advantage of the model in Lazarian \& Vishaniac (1999) is that, first of all, the reconnection is robust and is fast in any type of fluid, provided that the fluid is turbulent enough. The latter requirement is natural for most of astrophysical fluids (see Armstrong et al. 1995). At the same time, the model predicts that if the fluids are not turbulent initially, they should be prone to bursts of reconnection, which may provide
an appealing explanation of Solar Flares.

\subsection{Self-consistency of the model}

The high speed of reconnection given by equation (\ref{main})
naturally leads to a question of self-consistency.  Is it reasonable
to take the turbulent cascade suggested in GS95
when field lines in adjacent eddies are capable of reconnecting?
It turns out that in this context, our estimate for $V_{rec,global}$
is just fast enough to be interesting.  We note that when considering the 
intersection of nearly
parallel field lines in adjacent eddies the acceleration of plasma
from the reconnection layer due to the pressure gradient
is not $k_{\|}V_A^2$, but rather $(k_{\|}^3/k_{\perp}^2)V_A^2$,
since only the energy of the component of the magnetic field 
which is not shared is available to drive the outflow.  On the
other hand, the characteristic length contraction of a given
field line due to reconnection between adjacent eddies is only
$k_{\|}/k_{\perp}^2$.  This gives an effective ejection rate 
of $k_{\|}V_A$.  Since the width of the diffusion layer over a 
length $k_{\|}^{-1}$ is just $k_{\perp}^{-1}$, we can replace 
equation (\ref{main}) with
$V_{rec,global}\approx V_A {k_{\|}\over k_{\perp}}$. 
The associated reconnection rate is just
\begin{equation}
\tau^{-1}_{reconnect}\sim V_A k_{\|}, 
\end{equation}
which in GS95 is just the nonlinear cascade rate on the scale
$k_{\|}^{-1}$.  However, this result is general and does not
involve assuming that GS95 is correct.  
As we discuss below, most of the energy
liberated in reconnection goes into motions on length
scales comparable to the dimensions of the reconnecting eddies,
so this energy release will not short circuit the energy cascade
described in GS95.  On the other hand, we can invert this argument
to see that reconnection can play an important role in 
preventing the buildup of unresolved knots in the magnetic field.
Such structures could play a major role in inhibiting the cascade
of energy to smaller scales, flattening the energy spectrum relative
to the predictions of GS95.  Our conclusion is that such structures 
will disappear as fast as they appear, supporting the notion that
they play a limited role in the dynamics of MHD turbulence.

Finally, we note that if the magnetic field structure is driven by
turbulence in another location, as when the footpoints of magnetic
arcades are stirred by turbulent motions, then we can evaluate its
effects in terms of the amplitude of field stochasticity and the
scaling of structure anisotropy with scale. The actual turbulence may
be balanced or imbalanced, have or not have ``polarization intermittency'' 
(see Beresnyak \& Lazarian 2006, 2008) the robust nature of the reconnection 
implies that the reconnection will be sensitive to the amplitude
of the induced field stochasticity, but not the details of the
turbulent mixing process. 

\subsection{Turbulent diffusivity and dynamo}

To enable sustainable dynamo action and, for example, generate 
a galactic magnetic field, it is necessary to reconnect
and rearrange magnetic flux on a scale similar to a galactic
disc thickness within roughly a galactic turnover time ($\sim 10^8$~years).
This implies that reconnection must occur at a 
substantial fraction of the Alfv\'en velocity.
The preceding arguments indicate that such reconnection velocities should
be attainable if we allow for a realistic magnetic field
structure, one that includes both random and regular fields.
This does solve one part of the problem of dynamo. The other part is related to 
magnetic helicity conservation (see Vishniac et al. 2003). 

Interestingly enough, high turbulent diffusivity is also required for advecting
heat in astrophysical plasmas, e.g. in clusters of galaxies (see Lazarian 2006). 
Turbulent reconnection enables eddies to transfer heat in the way similar
to the advection of heat by turbulence in unmagnetized fluid (see Cho et al. 2003).

\subsection{Dissipation of energy}

The usual assumption for energy dissipation in reconnection
is that some large fraction of the energy given up by the
magnetic field, in this case $\sim \rho V_A^2 L_x^3$, 
goes into heating the electrons.  This is
not the case here.  Only a fraction, $\sim 1/(k_{\|}L_x)$ of any
flux element is annihilated by Ohmic heating within the reconnection
zone.  Over the entire course of
the reconnection event the efficiency for electron heating
is no greater than
\begin{equation}
\epsilon_e\lesssim {\eta/\Delta\over V_{rec,global}}={V_{rec,local}\over V_{rec,global}}=
{1\over k_{\|}L_x}.
\end{equation}
or, with GS95 scaling substituted (see LV99)
\begin{equation}
\epsilon_e\lesssim \left({V_Al\over\eta}\right)^{-2/5}
\left({v_l\over v_A}\right)^{8/5}\left({l\over L_x}\right)^{4/5}.
\end{equation}
The electron heating within the current sheet will not be uniform,
due to the presence of turbulence, the intermittent presence of
reconnected flux,
and any collective effects we have neglected here.  To the extent
that these are important they will also lower the electron
heating efficiency by broadening the reconnection layer. 

\subsection{Summary}

The results above can be summarized as follows


$\bullet$ Recent successful numerical testing of the model in Lazarian \& Vishniac (1999), presented
in a companion paper by Kowal et al. (2009, this volume) increase the appeal of the model and stimulate 
studies of its consequencies.

$\bullet$ The aforementioned model of stochastic field reconnection provides justification for many of astrophysical simulations. If, on the contrary, the only reconnection model that works is the collisionless reconnection, this means that most of the numerical simulations, for instance, of interstellar medium are in error. Indeed, in the collisional environments the reconnection speed would be negligible, which cannot be achieved with the numerical simulations.

$\bullet$ The model of magnetic field reconnection in the presence of weak magnetic field stochasticity is a natural generalization of the Sweet-Parker model of reconnection. Its many consequencies include bursts of reconnection, efficient first order Fermi acceleration of energetic particles, efficient diffusion of magnetic field in turbulent plasmas etc. 

{\bf Acknowledgments}. Research by AL is supported by the NSF grant AST 0808118 and the NSF Center for Magnetic Self-Organization and by EV is supported by the National Science and Engineering Research Council of Canada.

\end{document}